\documentclass[preprint,showpacs,showkeys,aps]{revtex4-1}
\usepackage{graphicx}
\usepackage{dcolumn}
\usepackage{bm}%

\begin{document}

\title{Oblique MHD shocks: space-like and time-like}

\date{\today}

\author{Ritam Mallick} 
\email{ritam.mallick5@gmail.com}
\affiliation{Frankfurt Institute for Advanced Studies, 60438 Frankfurt am Main, Germany}
\author{Stefan Schramm}
\email{schramm@th.physik.uni-frankfurt.de}
\affiliation{Frankfurt Institute for Advanced Studies, 60438 Frankfurt am Main, Germany}

\begin{abstract}
Shock waves constitute discontinuities in matter which are relevant in studying the plasma behaviour in astrophysical scenarios and in 
heavy-ion collision. They can produce conical emission in relativistic collisions and are also thought 
to be the mechanism behind the acceleration of energetic particles in active galactic nuclei and gamma ray bursts. The 
shocks are mostly hydrodynamic shocks. In a magnetic background they become magnetohydrodynamic (MHD) shocks.
For that reason we study the space-like and time-like shock discontinuity in a magnetic 
plasma. The shocks induce a phase transition in the plasma, here assuming a transition from hadron to quarks. The MHD conservation conditions
are derived across the shock.
The conservation conditions are solved for downstream velocities and flow angles for given upstream variables.
The shock conditions are solved at different baryon densities. For the space-like shocks 
the anisotropy in the downstream velocity arises due to the magnetic field. The downstream velocity vector always points downward with 
respect to the shock normal. 
With the increase in density the anisotropy is somewhat reduced. The magnetic field has effectively no effect on time-like shocks.
The slight anisotropy in the downstream flow velocities is caused by the boosting that brings the quantities from the fluid frame to normal 
incidence (NI) frame.  
\end{abstract}



\maketitle

\section{Introduction} \label{introduction}

Shock fronts or shock waves occur when disturbances in a medium propagate faster than 
the local speed of sound. Shocks are characterized by nearly discontinuous changes in the characteristics
of the medium, like pressure, temperature, energy etc. The width of the discontinuity is quite infinitesimal in
comparison to the actual system, and therefore it is assumed to be a single propagating front.

The relativistic magneto-hydrodynamic (MHD) theory of shock waves is being applied to problems of
cosmology and relativistic heavy-ion collisions. 
In the field of cosmology the shock waves are usually collision-less MHD shocks and are thought to be mostly 
responsible for the acceleration of particles in a variety of astrophysical objects ranging from active galactic 
nuclei (AGN) to gamma ray bursts (GRB) (known as first-order Fermi acceleration or diffusive shock acceleration). 
This phenomenon occurs when the charged particles in the plasma interact with the magnetic fields in the shock layers, and are
repeatedly transported back and forth across the shock and thereby gaining energy. The generation of such 
collision-less shocks have been extensively studied in the literature \cite{colburn,kunkel}. The commonly observed 
Earth bow shock indicate that the velocity anisotropic distributions (VAD) of the particles play an important
role in the shock formation. Observation with the new generation of air Cherenkov TeV $\gamma$-ray telescopes such as HESS, MAGIC
and VERITAS \cite{hinton} have established the fact that AGNs, microquasars and pulsar wind nebulae are 
powerful sources of high-energy photon radiation. They generate relativistic jets of particles, which collide with
the surrounding intergalactic and interstellar medium to give rise to high energy TeV $\gamma$ radiation. 
This is the currently accepted hypothesis. However, how the collision-less shock is generated 
is still not well understood.

Shock waves are usually created under rapid compression of matter. However, a similar discontinuity in matter 
can also be generated if the system suddenly expands and there is a phase transition (PT) \cite{taub}. Taub was the 
earliest one to study the relativistic hydrodynamic shocks, by writing the conservation equation across the shock
boundary also called Rankine-Hugoniot equation. De Hoffmann and Teller were first to study the magnetized
shock \cite{hoffmann}. This led the way for the study of MHD shocks for the following years \cite{landau,cabannes}. 
Lichnerowicz extended the analysis to the relativistic case \cite{lichnero}, where he treated the matter 
to be an ideal fluid with an infinite conductivity. The more recent important studies of MHD shocks \cite{majorana,ballard}
used relativistic equation of state to obtain the solutions from the conservation condition. The conservation 
condition of MHD shocks in a gyrotropic fluid has also been extensively studied \cite{gerbig}. Furthermore, several studies of MHD shocks,
the first-order Fermi acceleration and its connection with the astrophysical observation have also been performed \cite{kirk,ballard,summerlin}.

In the above mentioned astrophysical scenarios, the shock wave propagates with a velocity less than the velocity of light. 
That is, the normal vector of the surface of the discontinuity, is space-like. However, it may not always be the case. 
In some situations there may be a fast PT (first order) where the normal vector to the surface of the discontinuity 
can be time-like \cite{csernai}. A system undergoing rapid and homogeneous rarefaction,
bubbles at different spatial points are formed which are causally unconnected to each other. For such cases the phase 
boundary separating the two phases of matter becomes time-like. If the thickness of the time-like surface depends on the rate 
of formation of the bubbles and its growth, and if it is sufficiently thin, we can assume that the PT
happens along an approximate structureless time-like surface. An example of such type of PT is the 
hadronization of a supercooled quark-gluon plasma (QGP) in a heavy-ion collision. As the QGP fireball expands, it cools, and well below 
critical temperature the QGP hadronizes. The time-like shock hadronization in connection with heavy-ion collision has been studied 
extensively \cite{gorenstein,glendenning,csernai2}. The inflationary universe model can be thought to be a cosmological example.

The main aim of this work is to study both the space-like and time-like shock in a relativistic MHD formalism. In our specific 
calculation, we assume that
the shock converts hadronic matter to quark matter, and so the shock has on one side hadronic matter and on the other side 
quark matter. We treat the shock as a generalised oblique shock both in the De Hoffmann Teller frame and normal incidence 
frame. Such shock studies can be important both to the study of shocks in astrophysics and in heavy-ion collisions. In the 
astrophysical scenario such shock and PT can occur in neutron stars which have huge inner magnetic fields.
On the other hand, in the field of heavy-ion collisions such scenario can also occur when extremely large magnetic field 
are generated by colliding particles. In the next section we write the conservation equations for the time-like and space-like
shock front in the De Hoffmann frame. In section III we give the transformation equation from the fluid frame to the normal incidence frame. 
In section IV we present our results. In the final section (section V) we discuss our results and their applicability. 

\section{conservation condition for oblique shocks}

We denote the surface of discontinuity as $\Xi$ and the normal to the surface as $\Lambda^{\nu}$. The normalization condition 
is given as 
\begin{eqnarray}
 \Lambda^{\nu}\Lambda_{\nu}= +1  \: for\:time-like\:\Xi \\ \nonumber
                             -1  \: for\:space-like\:\Xi.
\end{eqnarray}

The energy momentum tensor of the system reads
\begin{equation}
 T^{\mu\nu}=wu^{\mu}u^{\nu}-pg^{\mu\nu},
\end{equation}
where, $w$ is the enthalpy ($w=e+p$), $e$ being the energy density and $p$ being the pressure. $u^{\mu}=(\gamma,\gamma v)$ is the 
4-velocity of the fluid, and is normalized such that $u^{\mu}u_{\mu}=1$ and $\gamma$ is the Lorentz factor. $g^{\mu\nu}$ is the 
metric tensor and is $(1,-1,-1,-1)$ using the flat space-time convention. The conservation conditions are nothing but the energy-momentum and baryon
number conservation laws, across the discontinuity of the shock surface. We denote $``1''$ as the initial state ahead of the shock 
front and $``2''$ as the final state behind the shock. The general derivation of the shock wave can be found in
\cite{csernai} and also in many subsequent literature \cite{rosenhauer,gorenstein}. One of the other important constraints of the 
transition is the thermodynamic stability condition, which requires non-decreasing entropy \cite{gyulassy}
\begin{equation}
 s_1u^{\mu}_1\Lambda_{\mu} \geq  s_2u^{\mu}_2\Lambda_{\mu},
\end{equation}
where, $s_1$ and $s_2$ are the entropy densities before and after the shock.

The generalised conservation conditions can be written as
\begin{eqnarray}
 T_1^{\mu\nu}\Lambda_{\nu}=T_2^{\mu\nu}\Lambda_{\nu} \\
 n_1u_1^{\mu}\Lambda_{\mu}=n_2u_2^{\mu}\Lambda_{\mu}.
\end{eqnarray}

The relativistic conservation conditions for the space-like (SL) and time-like (TL) shocks are derived from the above generalised equations.
Closely following ref. \cite{taub,landau,gorenstein}, the equation reads 

a. Space-like
\begin{eqnarray}
 T_1^{01}= T_2^{01} \\ \nonumber
 \Rightarrow w_1\gamma_1^2v_1=w_2\gamma_2^2v_2 ,\\
  T_1^{11}= T_2^{11} \\ \nonumber
 \Rightarrow w_1\gamma_1^2v_1^2+p_1=w_2\gamma_2^2v_2^2+p_2 ,\\
 n_1u_1^1=n_2u_2^1 \\ \nonumber
 \Rightarrow n_1v_1\gamma_1=n_2v_2\gamma_2
\end{eqnarray}

Solving the equations we get the solution of the downstream and upstream velocities, which are 
\begin{eqnarray}
 v_1^2=\frac{(p_2-p_1)(e_2+p_1)}{(e_2-e_1)(e_1+p_2)} \\
 v_2^2=\frac{(p_2-p_1)(e_1+p_2)}{(e_2-e_1)(e_2+p_1)}.
\end{eqnarray}

b. Time-like
\begin{eqnarray}
 T_1^{00}= T_2^{00} \\ \nonumber
 \Rightarrow w_1\gamma_1^2-p_2=w_2\gamma_2^2-p_2 ,\\
  T_1^{10}= T_2^{10} \\ \nonumber
 \Rightarrow w_1\gamma_1^2v_1=w_2\gamma_2^2v_2 ,\\
 n_1u_1^0=n_2u_2^0 \\ \nonumber
 \Rightarrow n_1\gamma_1=n_2\gamma_2
\end{eqnarray}

The corresponding downstream and upstream velocities are 
\begin{eqnarray}
 v_1^2=\frac{(e_2-e_1)(e_1+p_2)}{(p_2-p_1)(e_2+p_1)} \\
 v_2^2=\frac{(e_2-e_1)(e_2+p_1)}{(p_2-p_1)(e_1+p_2)}.
\end{eqnarray}

Comparing the matter velocities for the TL and SL shocks, we arrive at the equation
\begin{eqnarray}
 v_{1t}^2=\frac{1}{v_{1s}^2} \\ 
 v_{2t}^2=\frac{1}{v_{2s}^2}.
\end{eqnarray}
where, $t$ stands for TL shocks and $s$ for SL shocks, respectively.

Next, we address the conservation equation for the MHD shocks. In this case, the energy-momentum tensor has both matter and magnetic contributions.
We assume an ideal infinitely conducting fluid, therefore the electric field vanishes. The total energy momentum tensor is given by
\begin{equation}
 T_{\mu\nu}=T_{\mu\nu}^M+T_{\mu\nu}^B.
\end{equation}
The $M$ represents the matter part of the tensor and $B$ the magnetic part.
The magnetic part of the tensor is defined as
\begin{eqnarray}
 T_{00}^B=\frac{B^2}{8\pi} \\ 
 T_{ij}^B=\frac{B^2}{8\pi}\delta_{ij}-\frac{B_iB_j}{4\pi}.
\end{eqnarray}
where, $B_i$ is the magnetic field vector.

\begin{figure}
\vskip 0.2in
\centering
\begin{tabular}{cc}
\begin{minipage}{230pt}
\includegraphics[width=230pt]{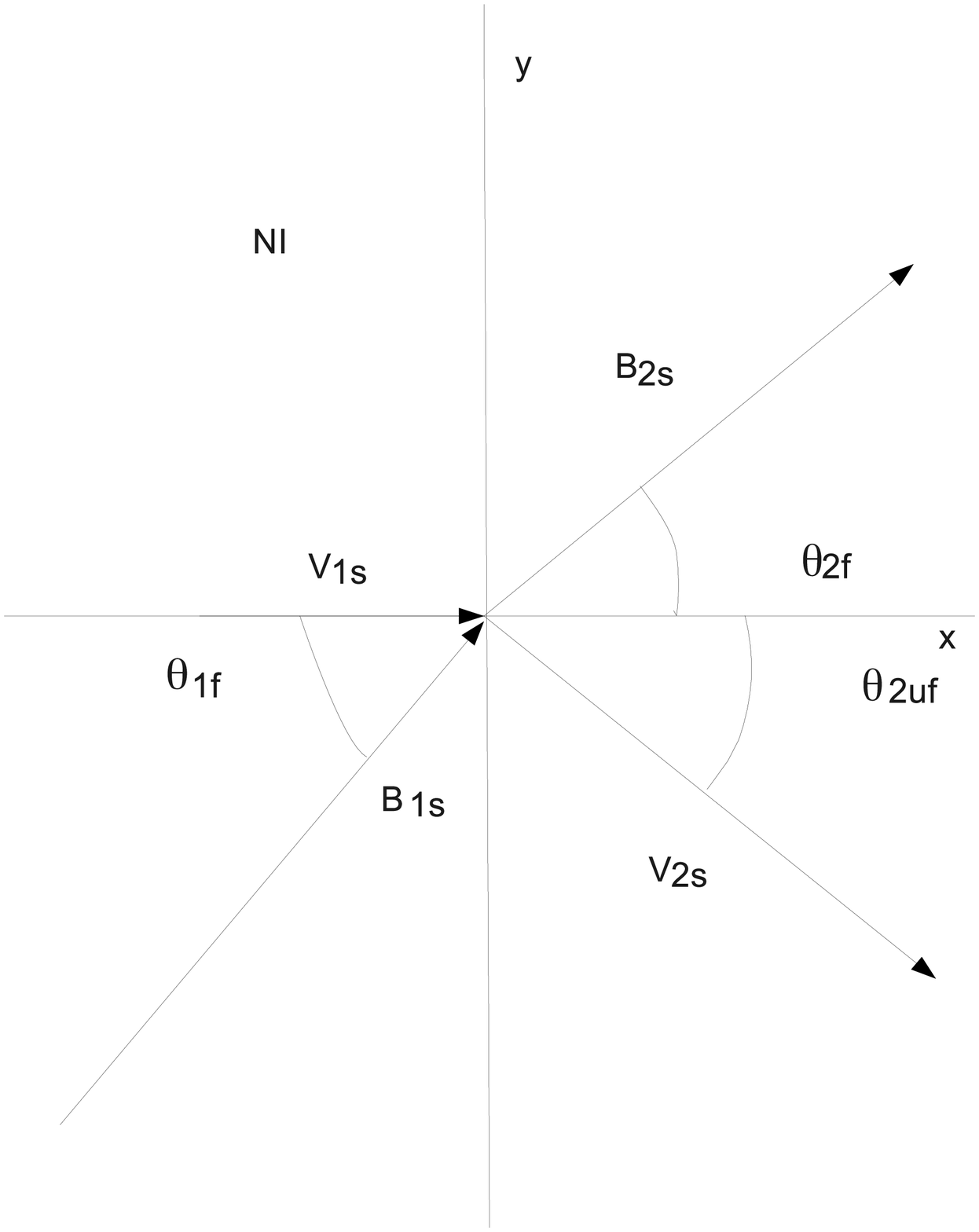} 
\end{minipage}
\begin{minipage}{230pt}
\includegraphics[width=230pt]{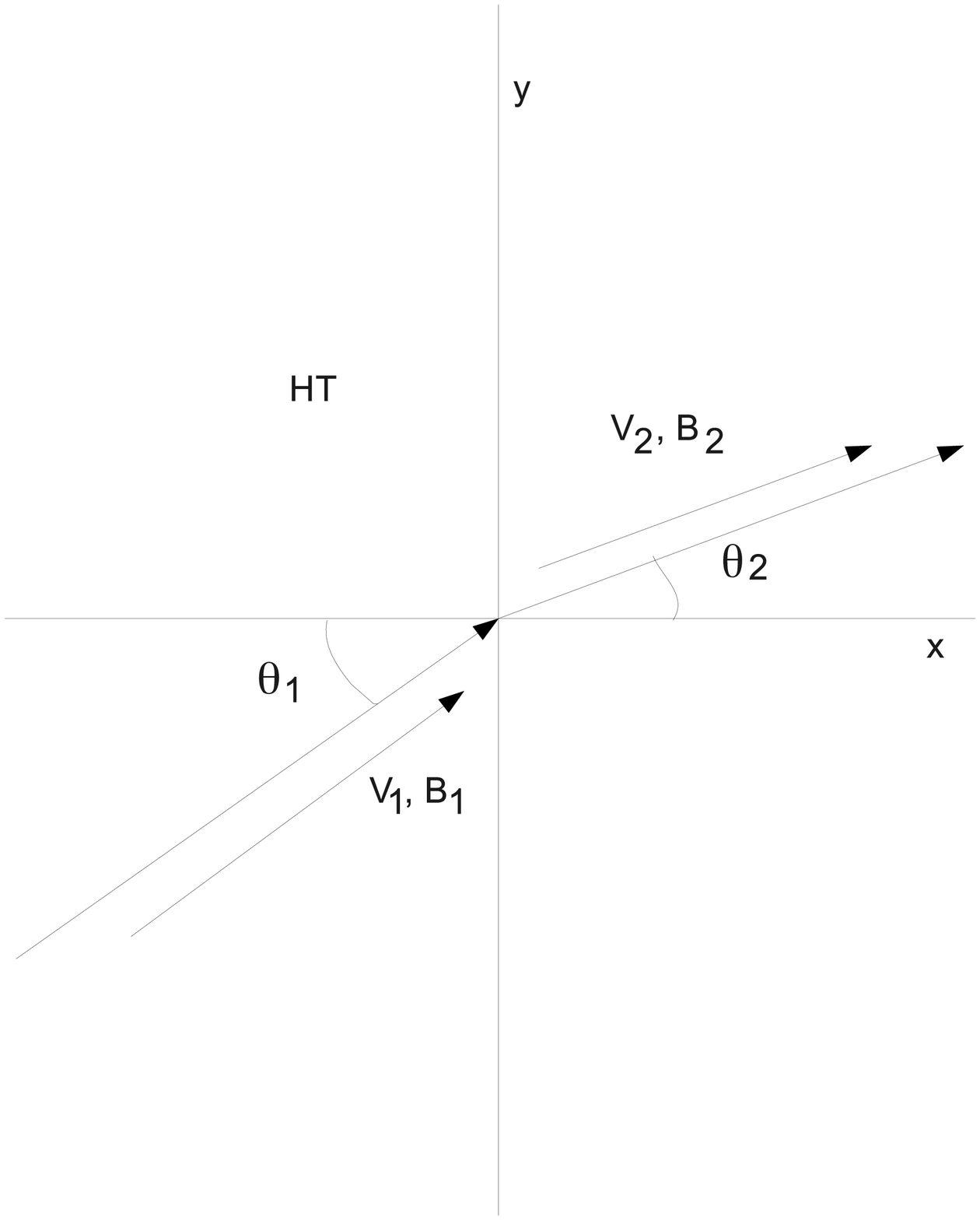} 
\end{minipage}
\\
\end{tabular}
\caption{Pictorial description of the upstream and downstream parameters for the HT and NI frame, respectively.}
\label{fig1}
\end{figure}

The shock conservation condition or the relativistic Rankine-Hugoniot condition is very difficult to solve for the MHD shocks.
To make it more manageable we go to the De Hoffmann Teller frame (HT) \cite{hoffmann}. The HT frame is the shock rest frame,
where there is no $\overrightarrow{u}\times \overrightarrow{B}$ drift electric fields. Therefore, for the subluminal flows the
HT frame is an obvious choice where the conservation conditions are reduced to simple forms. The system of equations are then transformed
from the HT frame to normal incidence (NI) frame, with a boost. The NI frame is a frame where the incident velocity is normal to the shock front.
Fig. \ref{fig1} gives the pictorial description of the two frames.
We then have a system of simple simultaneous equations in which the imaginary terms in the 
superluminal shocks are absent. Here we should mention that our approach of this frame transformation is similar to the one used by 
Ballard \& Heavens and Summerlin \& Baring \cite{ballard,summerlin}.

In this work we adopt the following conventions: The HT frame quantities do not have any subscript, whereas the 
NI frame quantities are labelled by a subscript $s$. The angle between the magnetic field and the shock normal 
in the HT frame is denoted by $\theta$ and that in the 
NI frame by $\theta_s$. Correspondingly, the angle between flow velocities and the shock normal is denoted by $\theta_u$ for the HT frame
and by $\theta _{us}$ in the NI frame. Along with the system of conservation equations we have also a set of equations of state (EOS) describing the 
matter phases in the upstream and downstream. We also assume that in the HT frame the fluid flows along the magnetic lines and there
is no $\overrightarrow{u}\times \overrightarrow{B}$ electric fields. The four matter jump conditions are given by the conservation of
baryon number, momentum (2 components) and energy density across the front. The electromagnetic jump condition is given by the 
equation
\begin{eqnarray}
 \nabla \cdot \overrightarrow{B}=0, \\
 \nabla \times \overrightarrow{E}=0.
\end{eqnarray}
The last term arises trivially because $\overrightarrow{E}=0$ holds everywhere.

Thus we define the oblique shock conservation condition. First we write the conservation condition in the HT frame.
In this frame the magnetic field and the matter velocities are aligned. Let us assume that the $x$-direction defines the normal 
to the shock plane. The magnetic field is constant and lies in the $x-y$ plane. Therefore the velocities in the $x$ and $y$ direction 
are given by $v_x$ and $v_y$, respectively. Similarly the magnetic fields in the $x$ and $y$ direction are given by $B_x$ and $B_y$.
The Lorentz factor is defined as 
\begin{equation}
\gamma=(1-v_x^2-v_y^2)^{-\frac{1}{2}}.
\end{equation}

With that the conservation conditions follow.

a. Space-like
\begin{eqnarray}
 T_1^{01}= T_2^{01} \\ \nonumber
 \Rightarrow w_1\gamma_1^2v_{1x}=w_2\gamma_2^2v_{2x},\\
  T_1^{11}= T_2^{11} \\ \nonumber
 \Rightarrow w_1\gamma_1^2v_{1x}^2+p_1+\frac{B_{1y}^2}{8\pi}=w_2\gamma_2^2v_{2x}^2+p_2+\frac{B_{2y}^2}{8\pi},\\
 T_1^{21}= T_2^{21} \\ \nonumber
 \Rightarrow w_1\gamma_1^2v_{1x}v_{1y}-\frac{B_{1x}B_{1y}}{4\pi}=w_2\gamma_2^2v_{2x}v_{2y}-\frac{B_{2x}B_{2y}}{4\pi},\\
 n_1u_1^1=n_2u_2^1 \\ \nonumber
 \Rightarrow n_1v_{1x}\gamma_1=n_2v_{2x}\gamma_2.
\end{eqnarray}

b. Time-like
\begin{eqnarray}
 T_1^{00}= T_2^{00} \\ \nonumber
 \Rightarrow w_1\gamma_1^2-p_1+\frac{B_{1y}^2}{8\pi}=w_2\gamma_2^2-p_2+\frac{B_{2y}^2}{8\pi},\\
 T_1^{10}= T_2^{10} \\ \nonumber
 \Rightarrow w_1\gamma_1^2v_{1x}=w_2\gamma_2^2v_{2x},\\
 T_1^{20}= T_2^{20} \\ \nonumber
 \Rightarrow w_1\gamma_1^2v_{1y}=w_2\gamma_2^2v_{2y},\\
 n_1u_1^0=n_2u_2^0 \\ \nonumber
 \Rightarrow n_1\gamma_1=n_2\gamma_2.
\end{eqnarray}

For the HT frame we also have 
\begin{eqnarray}
 \frac{v_{1y}}{v_{1x}}=\frac{B_{1y}}{B_{1x}}\equiv \tan\theta_{1} \\
 \frac{v_{2y}}{v_{2x}}=\frac{B_{2y}}{B_{2x}}\equiv \tan\theta_{2}.
\end{eqnarray}
With the assumption of infinite conductivity, the electric field is zero.
The Maxwell equation $\nabla \cdot \overrightarrow{B}=0$, defines that there are no monopoles and so we have
\begin{equation}
 B_{1x}=B_{2x}.
\end{equation}

However, this equations are valid only for subluminal flows. For broader applicability (including superluminal flows), we now transform 
the equations to the NI frame.

\section{transformation to NI frame}

As done by Summelin \& Baring and Kirk \& Heavens \cite{summerlin,kirk}, we arrive from the local
fluid frame to the NI frame by a boost of $v_{xs}$ along the $x$ direction and from the NI frame to the HT frame by a boost
of $v_{y^\star}$ in the $y$ direction. The shock planes of the HT and NI frames are coincident.

In the NI frame, as the upstream velocity is normal incident to the shock front the $y$ component of the velocity is zero ($v_{1ys}=0$).
The transformation of the velocities can be written as
\begin{eqnarray}
 v_{1x}=\frac{v_{1xs}}{\Gamma_{y^\star}}, \\
 v_{1y}=v_{y^\star}, \\
 v_{2x}=\frac{v_{2xs}}{\Gamma_{y^\star}(1+v_{y^\star}v_{2ys})}, \\
 v_{2y}=\frac{v_{2ys}+v_{y^\star}}{\Gamma_{y^\star}(1+v_{y^\star}v_{2ys})}. 
\end{eqnarray}

The two parameters that
connect the quantities of the HT frame with the NI frame are defined as
\begin{eqnarray}
 v_{y^\star}=v_{1xs}\tan\theta_{1f}, \\
 \Gamma_{y^\star}=(1-v_{y^\star}^2)^{-\frac{1}{2}}. 
\end{eqnarray}

The angles of the NI and the HT frame are connected by the relation
\begin{eqnarray}
 \tan\theta_{1}=\Gamma_{y^\star} \tan\theta_{1f}, \\
 \tan\theta_{2}=\Gamma_{y^\star} \tan\theta_{2f}.
\end{eqnarray}

The relationship between the magnetic field component in the two frames are given by
\begin{eqnarray}
 B_{x}=\frac{B_{xs}}{\Gamma_{y^\star}}, \\
 B_{y}=B_{ys}.
\end{eqnarray}

\begin{figure}
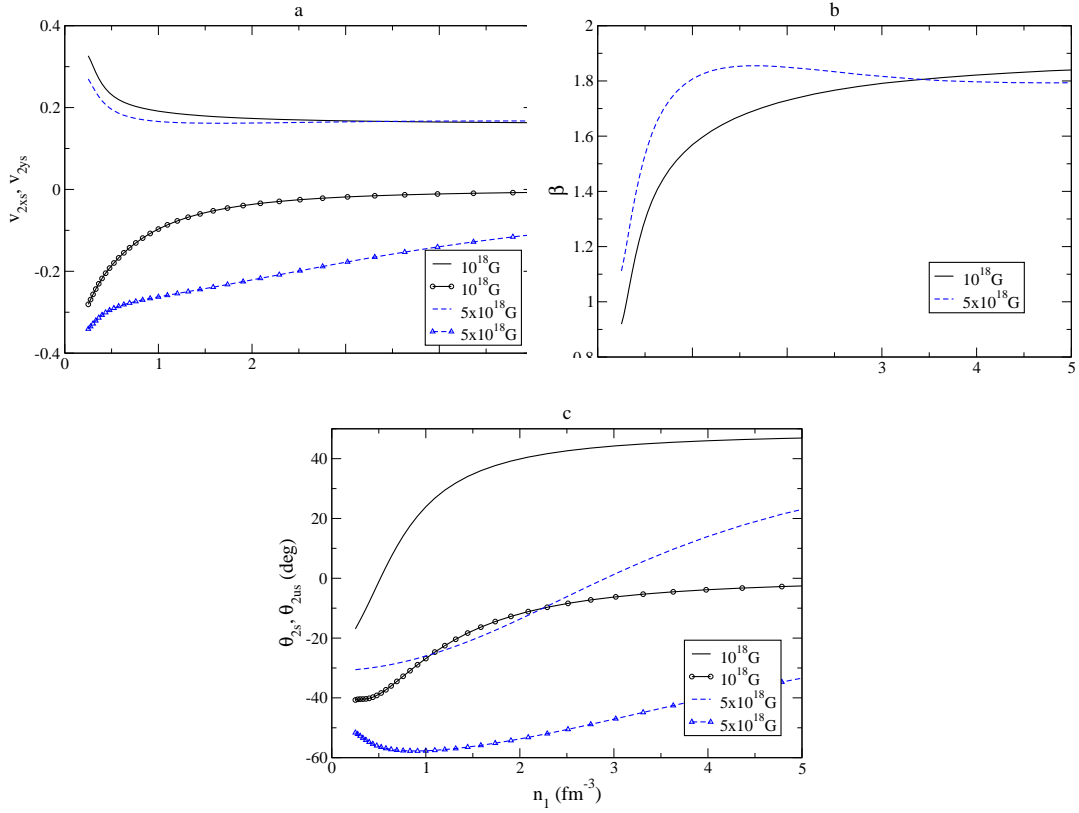

\vskip 0.2in
\centering
\begin{tabular}{cc}
\begin{minipage}{200pt}
\includegraphics[width=200pt]{fig3.eps} 
\end{minipage}
\begin{minipage}{200pt}
\includegraphics[width=200pt]{fig4.eps} 
\end{minipage}\\
\begin{minipage}{200pt}
\includegraphics[width=200pt]{fig5.eps} 
\end{minipage}
\\
\end{tabular}
\caption{Curves for the SL shock are shown in the figure as a function of density $n_1$. 
$v_{2xs}$,$v_{2ys}$, $\beta$, $\theta_{2s}$ and $\theta_{2us}$ are shown 
for two different values of magnetic field ($10^{18}$G and $5\times10^{18}$G). The incident velocity is $0.3$ and the incident magnetic 
angle $\theta_{1s}$ is $30\,^{\circ}$. In Fig. 2a, the full lines are for the $x$ component of the velocity and the dotted for the $y$ component. 
In Fig 2c, the full curves are used for downstream magnetic angle and the dotted curves are for flow velocities. We follow this convention
throughout the paper.}
\label{fig2}
\end{figure}

Now we need the relations transform brings the field angles from the fluid frame to the NI frame. These equations are given by
\begin{eqnarray}
 \tan \theta_{1f}=\Gamma_{1f}\tan \theta_{1s}, \\
 \tan \theta_{2f}=\Gamma_{2f}\tan \theta_{2s},
\end{eqnarray}
where, $\Gamma_{1f}=(1-v_{1xs}^2)^{-\frac{1}{2}}$ and $\Gamma_{2f}=(1-v_{2xs}^2)^{-\frac{1}{2}}$ respectively. The downstream 
flow angle is the same for fluid and NI frame ($\theta_{2uf}=\theta_{2us}$).

We solve the four conservation conditions in the HT frame. Subsequently, we transform the solution to the NI frame
and obtain the unknown solutions in this frame. The input to the equations are $p_1,n_1,v_{1xs}, \tan\theta_{1s}$ and $B_{1s}$.
The solution of the set of equations defines $p_2,n_2,v_{2xs},v_{2ys}$. We also have both the upstream and downstream EOS that 
relate $p$ and $e$ in the two phases. The importance of this methodology lies in the fact that the equation now has 
thermodynamic quantities relating the upstream and downstream components and the transformation velocities to get to the NI frame 
from the fluid frame. This method is useful for the fact that it removes all the imaginary and unphysical quantities in the 
superluminal shock and there is a smooth mathematical transition of the shock from the subluminal to the superluminal regime.

\section{Results}

\begin{figure}
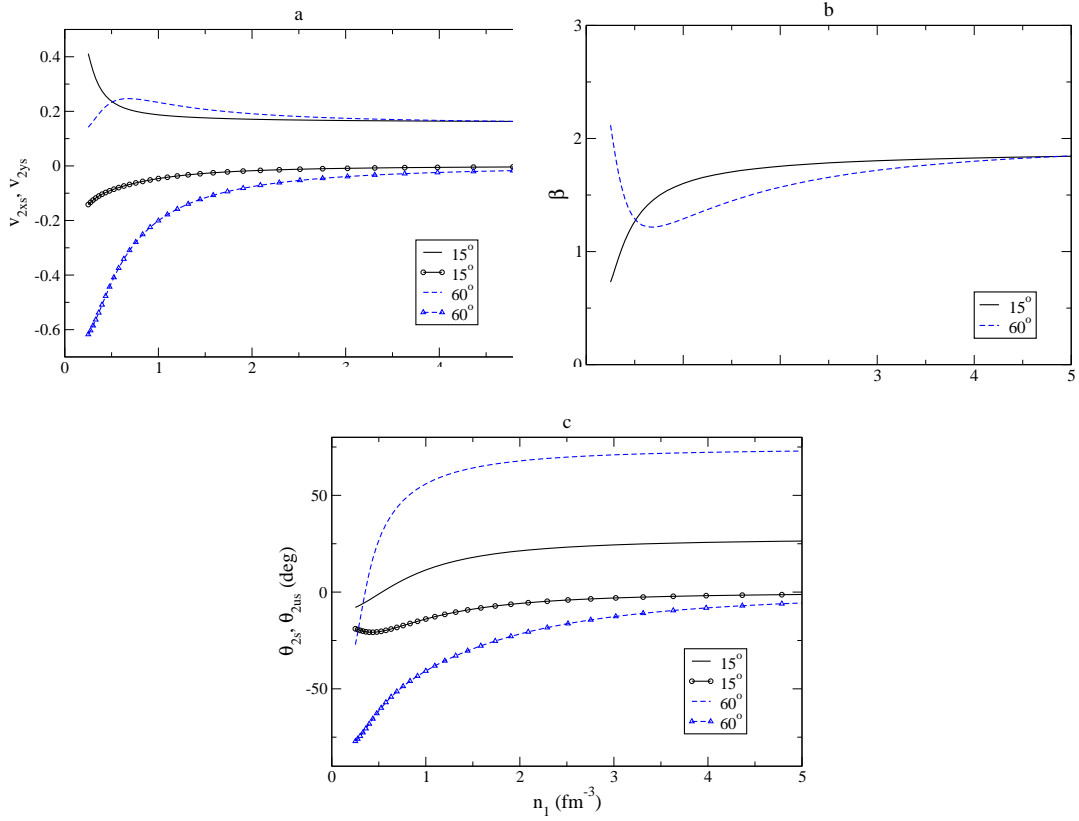

\vskip 0.2in
\centering
\begin{tabular}{cc}
\begin{minipage}{200pt}
\includegraphics[width=200pt]{fig6.eps} 
\end{minipage}
\begin{minipage}{200pt}
\includegraphics[width=200pt]{fig7.eps} 
\end{minipage}\\
\begin{minipage}{200pt}
\includegraphics[width=200pt]{fig8.eps} 
\end{minipage}
\\
\end{tabular}
\caption{$v_{2xs}$,$v_{2ys}$, $\beta$, $\theta_{2s}$ and $\theta_{2us}$ are plotted against density $n_1$ for two different upstream magnetic angles
($15\,^{\circ}$ and $60\,^{\circ}$). The incident velocity is $0.3$ and the magnetic field strength is $10^{18}$G. This 
figure shows curves for SL shocks.}
\label{fig3}
\end{figure}

In this work we have assumed constant magnetic field at all densities. Concentrating on the scenario of a hadron-to-quark PT, 
we have used a hadronic nonlinear Walecka model EOS \cite{walecka} 
to describe the upstream quantities and MIT quark bag model EOS \cite{mit} to describe the downstream quantities, respectively. 
The two EOS are standard choices and their detailed description can be found in previous literature. We assume that the shock front
induces a PT from normal hadronic mater to quark matter.

The input quantities are the upstream pressure, baryon number density, the magnetic field and the angle between the magnetic field 
and the shock normal. The effect of the magnetic field is quite negligible if the field strength is less than $10^{17}$G.
If it is larger it affects both downstream velocities and the downstream magnetic and flow angles. First we analyse the 
SL shock. For a fixed upstream velocity ($0.3$ in units of $c$) and a fixed magnetic angle ($\theta_{1s}=30\,^{\circ}$), as we increase the 
magnetic field, the downstream $x$ component of the velocity decreases slightly (usually at low densities, Fig. \ref{fig2}). However, the $y$ 
component of the downstream velocity increases with increasing magnetic field, but the sign is negative, which means that the angle 
between the downstream flow velocity and the shock normal is negative. The positive and negative directions/angles convention are chosen with 
respect to the angle between the incident velocity and the incident magnetic field (shown in Fig. \ref{fig1}). 
We define a quantity $\beta=\frac{v_{1xs}}{v_{2xs}}$, which is the velocity 
compression ratio. at low density $\beta$ increases with increase in magnetic field. The magnetic field acts as an extra pressure term which enhances
the compression of the medium. Next we show the downstream magnetic and flow angles. As can be seen in Fig. \ref{fig2}b, $\beta$ is greater than $1$
(other that at very low densities). This shows that we have velocity compression due to the PT. At higher densities $\beta$
saturates and is quite independent of the magnetic field strength. As the $y$ component of the velocity is negative the flow 
angle is also negative, that is, the downstream velocity vector points downwards with respect to the shock normal. 
The flow angle becomes more negative with 
increasing magnetic field, which points to the fact that the velocity anisotropy in the downstream flow is directly proportional to the magnetic field.
The downstream magnetic angle is also slightly negative and becomes more negative with increasing magnetic field strength.

In the plots (Fig. \ref{fig2}) we have also shown shocks occurring at different densities. However, we should not confuse them 
to be some kink of temporal evolution. As the density increases the flow velocity decreases (both $x$ and 
$y$ component). This is because as matter gets denser, it generates a much higher damping effect for the velocities. 
The parameter $\beta$ increases with density, because for the same value of incident velocity the downstream velocity reduces 
with increasing density. The flow angle first increases because the $x$ 
component of the downstream velocity decreases much faster than the $y$ component, but after some density ($4-6$ times saturation density) 
it decreases, because from there on $y$ component decreases much faster than the $x$ component. With increasing density the 
magnetic angle tends to move from negative to positive values.

\begin{figure}
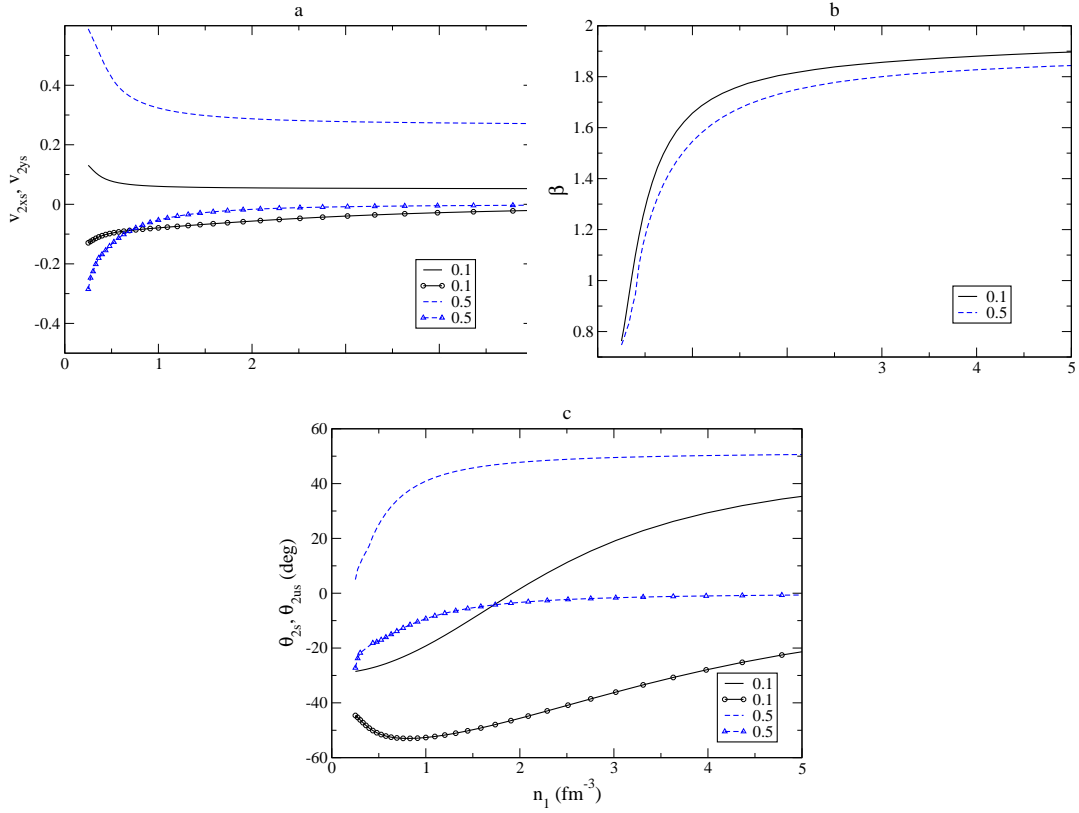

\vskip 0.2in
\centering
\begin{tabular}{cc}
\begin{minipage}{200pt}
\includegraphics[width=200pt]{fig9.eps} 
\end{minipage}
\begin{minipage}{200pt}
\includegraphics[width=200pt]{fig10.eps} 
\end{minipage}\\
\begin{minipage}{200pt}
\includegraphics[width=200pt]{fig11.eps} 
\end{minipage}
\\
\end{tabular}
\caption{The SL shock parameters $v_{2xs}$,$v_{2ys}$, $\beta$, $\theta_{2s}$ and $\theta_{2us}$ are plotted against density $n_1$ for two different 
incident velocities ($0.1$ and $0.5$). The incident magnetic angle is $30\,^{\circ}$ and the magnetic field strength is $10^{18}$G.}
\label{fig4}
\end{figure}

In Fig. \ref{fig3}, we plot curves for a fixed magnetic field  ($10^{18}$G) and for a fixed upstream velocity ($0.3$) with different
incident magnetic field inclination ($\theta_{1s}$). 
For smaller incident angle, the $x$ component starts with a high value ($0.4$) and then decreases with density, saturating at higher densities.
For larger incident angle, the behaviour is quite opposite, it starts with some low value ($0.16$) and increases with density. It attends a peak
($0.23$) and then decreases and saturates at higher densities. The saturation value is quite similar for a different incident magnetic angle. 
As, the $x$ component of the velocity shows such a behaviour, the $\beta$ also reflects its behaviour. For larger incident angle it first
decreases and then increases with density. However, the value is always greater than $1$.

The $y$ component of the velocity increases (becomes more negative) with increase in incident angle. 
The incident magnetic angle adds to the anisotropy 
of the downstream velocity. This can be seen from the flow angle, which increases with increase in
upstream magnetic angle. The downstream magnetic angle ($\theta_{2s}$) is directly proportional to the upstream magnetic angle 
($\theta _{1s}$). As the density 
increases the downstream magnetic angle increases, however, the flow angle decreases. 

Now keeping the magnetic field ($10^{18}$G) and incident angle fixed ($30\,^{\circ}$) (Fig. \ref{fig4}), if we increase the incident velocity the 
downstream $x$ component of the velocity increases and saturates at larger densities. The $y$ component of the downstream velocity for small
incident velocity is small and does not decrease much. However, for large incident velocity, initially $y$ is large, but decreases very fast and
saturates at a value very close to $0$. $\beta$ for smaller incident velocity is slightly larger than for larger incident velocity. $\beta$ 
increases very fast initially and saturates at higher densities.
As the incident velocity increases the anisotropy in the downstream velocity decreases. 
This can be seen by the fact that the flow angle decreases with increasing upstream velocity. The magnetic 
angle (which is positive) increases with the increase in upstream velocity. 

Overall, from the above set of curves, we find that the anisotropy in the downstream velocity is enhanced by the magnetic field and also to some extent by
the incident magnetic angle. The downstream velocity vector always points downward. Higher incident velocity tends to lessen this effect. With the 
increase in density the anisotropy is also reduced. $\beta$ is usually greater than $1$.

\begin{figure}
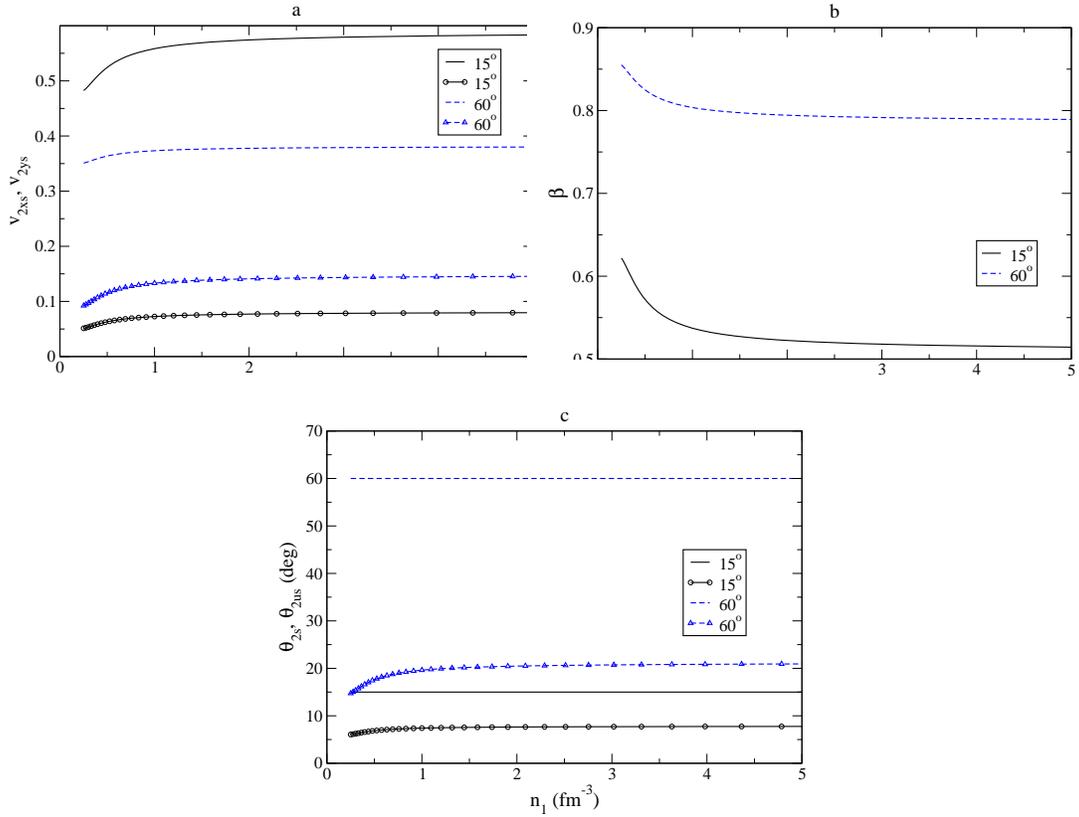

\vskip 0.2in
\centering
\begin{tabular}{cc}
\begin{minipage}{200pt}
\includegraphics[width=200pt]{fig12.eps} 
\end{minipage}
\begin{minipage}{200pt}
\includegraphics[width=200pt]{fig13.eps} 
\end{minipage}\\
\begin{minipage}{200pt}
\includegraphics[width=200pt]{fig14.eps} 
\end{minipage}
\\
\end{tabular}
\caption{Curves are plotted for TL shocks, where the behaviour of $v_{2xs}$,$v_{2ys}$, $\beta$, $\theta_{2s}$ and $\theta_{2us}$ 
are shown as function of density $n_1$, 
for two different incident magnetic angles ($15\,^{\circ}$ and $60\,^{\circ}$). The incident velocity is $0.3$.}
\label{fig5}
\end{figure}

\begin{figure}
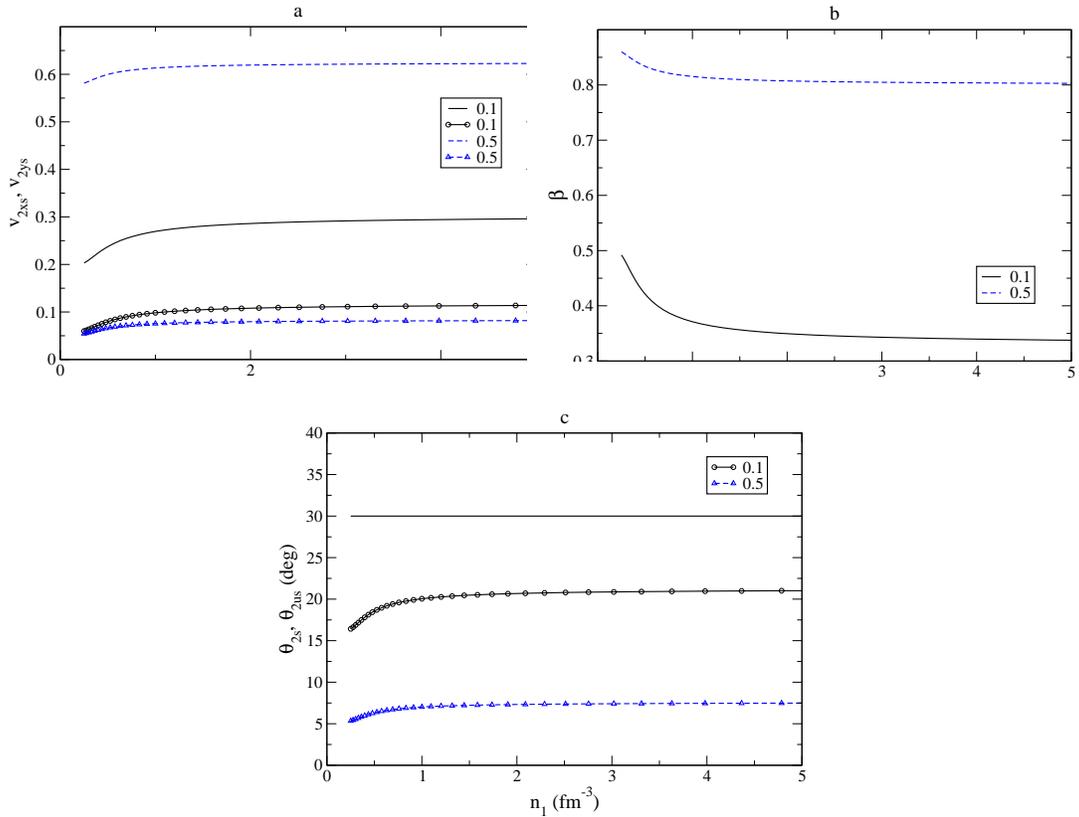

\vskip 0.2in
\centering
\begin{tabular}{cc}
\begin{minipage}{200pt}
\includegraphics[width=200pt]{fig15.eps} 
\end{minipage}
\begin{minipage}{200pt}
\includegraphics[width=200pt]{fig16.eps} 
\end{minipage}\\
\begin{minipage}{200pt}
\includegraphics[width=200pt]{fig17.eps} 
\end{minipage}
\\
\end{tabular}
\caption{$v_{2xs}$,$v_{2ys}$, $\beta$, $\theta_{2s}$ and $\theta_{2us}$ versus density $n_1$ for TL shocks are plotted
for two different incident velocities ($0.1$and $0.5$). The magnetic angle is $30\,^{\circ}$. As the magnetic angle is unaffected by 
the shock, in Fig \ref{fig6}c the black straight line marks the downstream magnetic angle, which is the same for two incident velocity.}
\label{fig6}
\end{figure}

For the TL shocks, the magnetic fields have no contribution to the shock velocities (Fig. \ref{fig5} and Fig. \ref{fig6}). 
The upstream and downstream magnetic 
angles remain the same, before and after the shock. The magnetic field has no effect on the downstream flow velocities or on the flow angle.
The anisotropy in the downstream flow velocities is caused by the boost that brings the quantities from the fluid frame to the NI frame. 
The $x$ component of the downstream velocity decreases with incident magnetic angle, whereas the $y$ component decreases (Fig. \ref{fig5}). 
For the TL shocks both the $x$ and $y$ component of the downstream velocities are positive. Therefore, an increase in the
incident magnetic angle makes the downstream velocity more anisotropic.
The TL shock differs from the SL shock by the fact that the $y$ component of the velocity 
is positive and so is the flow angle. The downstream velocity vector points upwards. This is the case because the anisotropy is generated by 
the boost and not 
by the magnetic field. The $\beta$ parameter also differs, it is always less than $1$ and decreases with density. 
This shows that there is no velocity compression but rarefaction in velocity due to the PT. 
As the magnetic 
field has no effect on the TL shock, the magnetic angle remains the same before and after the shock. The flow angle (positive for 
TL shocks) increases with incident magnetic angle as the anisotropy of the downstream velocity increases.

The $x$ component of the downstream velocity 
is directly proportional to the upstream velocity (Fig. \ref{fig6}), and has a small effect on the $y$ component. 
The velocity component $x$ in the downstream increases with increase in incident velocity whereas, the $y$ component decreases. $\beta$ 
increases with increase in incident velocity. The flow angle decreases with increase in incident velocity, and so the anisotropy in 
the downstream velocity is reduced.

The above discussion highlights that the system does not feel 
the magnetic effect, as if the fluid is non magnetic. Thus, the magnetic field present in a neutron star will not affect a 
TL shock, and so the PT would also remain unaffected. Even in the case of a heavy-ion collision the magnetic field would have 
no effect on a TL shock produced by the transition of a QGP to hadrons during the expansion of the fireball.
This is due to the fact that the electromagnetic field, which we assumed, 
that one can neglect the TL component (electric) compared to the SL component (magnetic). 

\section{Summary and discussion}

Summarising our work, we have derived the general Rankine-Hugoniot condition for a MHD oblique shock.
First we have derived the equations for the subluminal shock in the HT frame and from there we have derived a more general shock
(sub and superluminal shock) in the NI frame. To go from the fluid frame to the NI frame we have used a set of boosts along the 
coordinate axes. This is very useful as it removes all the imaginary and unphysical quantities present in the 
superluminal shock. 

In the HT frame we have written the four matter conservation equations and two electromagnetic conservation equations (Maxwell equations).
There is also a set of equations which transforms the quantities from the fluid to the NI frame. We have formulated the equations for 
both the SL and TL shock and the SL conservation condition matches well with previous works \cite{summerlin}. 
Along with these equations we also have introduced a set of standard EOS describing matter in 
the upstream and downstream phases. We assume that the shock wave brings about a PT from hadronic to quark matter.

The important results of this work are relevant both for astrophysics and heavy-ion collision.
The input quantities are the upstream variables and we are solving the conservation equation to obtain the downstream variables. 
The magnetic effects are relevant once the field strength is greater than $10^{17}$G. The general behaviour for SL shocks, 
the anisotropy in the downstream velocity is enhanced by the magnetic field and also to some extent by
the incident magnetic angle. Higher incident velocity and higher baryon density tends to reduce this effect. 
The TL shock differs from SL shocks, as the magnetic field has effectively no effect on the former.
The slight anisotropy in the downstream flow velocities is caused by the boosting that brings the quantities from the fluid frame to the NI frame. 
However, the cause may lie in our assumption of small electric fields. In the SL
shocks there is compression in the velocity ratio whereas in the TL shocks there is rarefaction in the velocity ratio due to PT.

In this work we consider a special case of PT from hadronic to quark matter. This is a phenomenon that might 
take place inside a neutron stars (NS), where the star is ultimately converted to a quark star (QS). Some perturbation (like 
spin-down \cite{glendenning1}) may induce such a PT brought about by a shock wave, originating at the centre. If such is the case, then
a detailed study of shock waves inside the star is needed. The first step towards it is the study of MHD shock waves in those environments.
Here we study such a scenario where a MHD shock wave (both SL and TL) brings about a PT from hadronic to quark matter.
We find that the SL and TL MHD shock waves are quite different from one another. The downstream flow 
velocity vectors for SL shocks points in opposite direction (i.e. the flow angle has opposite signs) in comparison to the 
TL shocks. The velocity compression ratio is also typically different. And the most significant difference comes from the magnetic 
effect. For the SL shocks all the downstream components depends strongly on the magnetic field and its angle with respect to the shock normal. 
Whereas, TL shocks are unaffected by magnetic field. The difference in the downstream behaviour of matter variables for SL and TL shocks 
could have some observational consequences for NS. The observed pulses from the NS are the jets of particles emitted by the NS and are directly in 
our line of sight. The jets of particles interacts with the NS environment surrounding it. Also, it was suggested
\cite{bombaci,berez,mallick} that gamma ray bursts (GRB) can be a consequence of PT in NS. Therefore, if the PT is brought about by a 
shock wave, its signature may differ for SL and TL shocks and could be strongly dependent on the magnetic field involved.

It is suggested that the particle acceleration and production of high energy (TeV) $\gamma$-rays from blazars and AGNs depends on the 
nature of the shock environment, the shock speed and the magnetic field present there. The spectral indices which are measured depends
strongly on these parameters, and provide deep insight regarding blazars and GRB's. This may be also true for shock waves in NS, where 
the downstream flow variables for SL and TL shocks are quite different. However, for more significant observational
inferences, we need more detailed calculation of the particle emission, their acceleration and their interaction with the NS environment.
In heavy-ion collision, the PT is quite the opposite. There we have hadronization of particles from an expanding and subsequently cooling 
QGP fireball.
The PT can also be thought to be initiated by a shock wave. Although, in our work we analyse the opposite PT, the results
can provide a insight even for heavy-ion collision, for the fact that the magnetic field which are produced in heavy-ion collisions 
would have no effect on the TL shock whereas it would significantly affect the SL shocks.

We should mention that the transformation equations from the fluid to the NI frame are similar to Ballard \& Heavens and Summerlin \& Baring 
(\cite{ballard,summerlin}). We have solved the Rankine-Hugoniot condition
for the MHD fluid with the assumption that there is a PT from hadronic to quark matter. There is no temporal evolution.
To look into the temporal evolution or how shock waves bring about a PT in neutron stars we have to study and solve the 
corresponding equations of motion (the continuity and Euler equation). However, before tackling such complex problems we can guess the 
initial conditions from this work. Therefore, these results can serve as the starting point of such calculations which we hope to report 
upon in the future work.

\begin{acknowledgments}
R. M. would like to thank HIC for FAIR for providing financial support.
\end{acknowledgments}

{}
\end{document}